\documentstyle[12pt,aaspp4]{article}

\begin{document}

\title{\bf Dynamics of Magnetic Flux Elements in the Solar Photosphere}

\author{A. A. van Ballegooijen, P. Nisenson, R. W. Noyes}
\affil{Harvard-Smithsonian Center for Astrophysics,
60 Garden Street, Cambridge, MA 02138}
\authoremail{vanballe@cfa.harvard.edu}
\author{M. G. L\"{o}fdahl}
\affil{Lockheed Martin Missiles and Space, Advanced Technology Center,
3251 Hanover Street, Palo Alto, CA 94304}
\authoremail{lofdahl@sag.space.lockheed.com}
\author{R. F. Stein}
\affil{Michigan State University, Dept. of Physics \& Astronomy,
East Lansing, MI 48824}
\authoremail{stein@msupa.pa.msu.edu}
\author{{\AA}. Nordlund}
\affil{Copenhagen University Observatory, {\O}stervoldgade 40,
DK-1350, Copenhagen, Denmark}
\authoremail{aake@tac.dk}
\and
\author{V. Krishnakumar}
\affil{Indian Institute of Astrophysics, Koramangala, Bangalore
560034, India}
\authoremail{krishna@iiap.ernet.in}

\begin{abstract}
The interaction of magnetic fields and convection is investigated
in the context of the coronal heating problem. We study the motions
of photospheric magnetic elements using a time series of high
resolution G-band and continuum filtergrams obtained at the Swedish
Vacuum Solar Telescope at La Palma. The G-band images show bright
points arranged in linear structures (``filigree'') which are located
in the lanes between neighboring granule cells. We measure the motions
of these bright points using an object tracking technique, and we
determine the autocorrelation function describing the temporal
variation of the bright point velocity. The correlation time of the
velocity is about 100 s. To understand the processes that determine
the spatial distribution of the bright points, we perform simulations
of horizontal motions of magnetic flux elements in response to solar
granulation flows. Models of the granulation flow are derived from the
observed granulation intensity images, using a simple 2D model that
includes inertia and horizontal temperature gradients; the magnetic
flux elements are assumed to be passively advected by this granulation
flow. The results suggest that this passive advection model is in
reasonable agreement with the observations, indicating that on a time
scale of 1 hour the flux tubes are not strongly affected by
their anchoring at large depth. Finally, we use potential-field
modeling to extrapolate the magnetic and velocity fields to larger
height. We find that the velocity in the chromosphere can be locally
enhanced at the separatrix surfaces between neighboring flux tubes.
The predicted velocities are several km/s, significantly larger than
those of the photospheric flux tubes. The implications of these
results for coronal heating are discussed.
\end{abstract}

\section{Introduction}

Observations of the Sun in X-rays show that the corona is heated to
several million degrees, and that magnetic fields play a key role in
this heating process (e.g. Vaiana \& Rosner 1978; Golub et al. 1980;
Pallavicini et al. 1981; Kano \& Tsuneta 1995, 1996; Shimizu \&
Tsuneta 1997; Yoshida \& Tsuneta 1996; Falconer et al. 1997). 
The energy source for this heating must lie in the turbulent
convection zone below the photosphere.
The interaction of the magnetic field with convective flows produces
two types of magnetic disturbances. First, the buffeting of magnetic
flux tubes by the granulation flow generates transverse MHD waves
(e.g. Steiner, Kn\"{o}lker \& Sch\"{u}ssler 1994), which propagate
upward along the magnetic flux tubes and dissipate their energy in the
chromosphere or corona (see Ofman, Klimchuk \& Davila 1998, and
references therein). Second, in coronal loops the random motions
of the footpoints produce twisting and braiding of coronal field
lines, which generates field-aligned DC electric currents that can
be dissipated resistively (Parker 1972, 1983; Tucker 1973; Rosner
et al. 1978; Sturrock \& Uchida 1981; Heyvaerts \& Priest 1984; van
Ballegooijen 1985, 1986, 1990a; Milano, Gomez \& Martens 1997). The
main difference between these processes is that plasma inertia plays a
key role in MHD wave propagation but is not important for the dynamics
field-aligned currents. Therefore, magnetic heating mechanisms
can be crudely classified as either wave-heating or current-heating
mechanisms (for reviews of coronal heating theory, see Narain \&
Ulmschneider 1990, 1996; Zirker 1993).

The relationship between coronal heating and the dynamics of
photospheric magnetic structures is not well understood.
Van Ballegooijen (1985, 1986) proposed that the slow twisting and
braiding of coronal field lines by horizontal footpoint motions leads
to a {\it cascade of magnetic energy} within coronal loops. The heating
rate predicted by this model is $\epsilon \sim B_0^2 D / (4 \pi L^2)$,
where $B_0$ is the coronal field strength, $L$ is the loop length, and
$D$ is the ``diffusion constant'' describing the random walk of
photospheric flux elements. The latter is given by $D \sim u_0^2
\tau_0$, where $u_0$ is the typical velocity of the magnetic elements,
and $\tau_0$ is the velocity coherence time. Observations of the
spreading of active regions indicate that $D$ is in the range 150
to 425 $\rm km^2 ~ s^{-1}$ (DeVore et al. 1985), but such values
of $D$ yield coronal heating rates which are a factor $\sim 40$ too
small compared to observed radiative and conductive loss rates
(van Ballegooijen 1986). We suggest there may exist short-period
motions with velocities of 1-2 km/s which contribute to coronal
heating but do {\it not} contribute to the spreading of active region
flux on time scales of days to months (i.e., the motions cannot be
described by a ``random walk''). Short-period motions of magnetic
elements have indeed been observed (e.g. Berger \& Title 1996),
but the characteristics of such motions are not well understood,
and it is unclear whether the velocities are sufficient to explain
the observed coronal heating.

A number of authors have developed three-dimensional simulation
models of the magnetic structure and heating of solar coronal loops
(e.g. van Ballegooijen 1988, 1990b; Mikic, Schnack \& Van Hoven 1989;
Longcope \& Sudan 1994; Hendrix et al. 1996; Galsgaard \& Nordlund
1996). In these simulations an initially uniform field connecting two
parallel plates is considered (the plates represent the photosphere
at the two ends of a coronal loop). The footpoints at the boundary
plates are subjected to a series of randomly phased, sinusoidal
flows patterns. The flows are incompressible, so that the component of
magnetic field perpendicular to each boundary plate remains uniform.
This is obviously not a good model of the actual conditions in the
solar photosphere. Observations indicate that the photospheric
magnetic field is highly intermittent, consisting of discrete flux
elements with nearly field-free gas in between (e.g. Title, Tarbell
\& Topka 1987; Title et al. 1992). The solar convection plays a key
role in producing these flux elements, which are mainly located
at the boundaries between granulation cells. This concentration of
flux into discrete elements probably has important consequences for
coronal heating: if the magnetic field consists of a bundle of
topologically distinct flux tubes, then the random motions of these
flux tubes in the photosphere will create {\it tangential
discontinuities} at the interfaces between the tubes in the corona
(Parker 1972; Glencross 1975, 1980; Rosner et al. 1978; Sturrock \&
Uchida 1981; D\'{e}moulin \& Priest 1997). Clearly, to understand
where and how the corona is heated, it is necessary to study the
dynamics of photosheric flux elements and to develop coronal heating
models which take the presence of such flux elements into account.

In this paper we focus on the first part of this problem, namely,
the dynamics of magnetic flux elements in the photosphere and
chromosphere. We analyze observations of G-band bright points obtained
at the Swedish Vacuum Solar Telescope (SVST) on La Palma, and derive
the autocorrelation function describing the temporal variations
of the bright point velocity. We find that the velocity changes on a
time scale of about 100 s. We develop an empirical model of the
granulation flow and simulate the horizontal motions of flux tubes,
assuming they are passively advected by the granulation flow.
By comparing the results of these simulations with the observed
spatial distributions and velocities of G-band bright points,
we show that this passive advection model is in reasonable agreement
with the observations. Finally, we use potential-field modeling
to extrapolate the magnetic and velocity fields to larger heights
(up to 1500 km in the chromosphere). The results indicate that
the spreading of the flux tubes with height and their interactions
with each other produce plasma flows in the chromosphere with
velocities of several km/s, much faster than the velocities of
the underlying photospheric flux tubes. This suggests that the
coronal heating rate can be significantly enhanced by the
three-dimensional geometry of the flux tubes in the photosphere
and chromosphere.

\section{Summary of Bright Point Observations}

Observations of the Sun with high spatial resolution show network
bright points (Muller 1983, 1985, 1994; Muller \& Keil 1983;
Muller \& Roudier 1984, 1992) and ``filigree'' (Dunn \& Zirker 1973;
Mehltretter 1974; Berger et al. 1995), which are small bright
features located within the intergranular lanes. The bright points
and filigree are seen  in the wings of strong spectral lines such
as H$\alpha$ and Ca II H \& K, in lines formed in the photosphere, 
and even at continuum wavelengths (with reduced contrast). The widths
of these structures is 100 to 200 km, at the limit of resolution of
ground-based solar telescopes.
The bright points are associated with regions of strong
magnetic field (Chapman \& Sheeley 1968; Title, Tarbell \& Topka 1987;
Simon et al. 1988; Title et al. 1989, 1992; Keller 1992) and
correspond to magnetic flux tubes of kilogauss field strength that
stand nearly vertically in the solar atmosphere (Stenflo 1973;
Stenflo \& Harvey 1985; Sanchez Almeida \& Martinez Pillet 1994; see
review by Solanki 1993). The granules near network bright points are
smaller and more numerous than near a normal intergranular space
(Muller, Roudier \& Hulot 1989). The filigree produce abnormal
granulation patterns (Dunn \& Zirker 1973) and appear to be chains of
bright points which fill the intergranular lanes.

The dynamical behaviour of bright points has been studied by a number
of authors. Muller (1983) found that facular points on the quiet sun
are predominantly located in patches at the periphery of supergranule
cells, indicating that the magnetic elements are advected by the
supergranular flow. The bright points always first appear in the dark
spaces {\it at the junction of several granules}, never inside a
granule nor in the space between only two granules. As the granulation
pattern evolves, the bright points remain in the intergranular spaces
throughout their lifetime, but not necessarily at the junction of
several granules like at the time of their first appearance. New
bright points have a tendency to appear adjacent to existing
points, and 15 \% of the bright points seem to split into two points
which move apart until a separation of 1 to 1.5 arcsec is reached.
Muller et al. (1994) measured velocities of 29 isolated bright points
and found a mean speed of 1.4 km/s. Strous (1994) studied bright points
in a growing active region. Using line-center images taken in Fe I 5576
{\AA}, he found velocities between 0.26 km/s and 0.62 km/s. Berger \&
Title (1996) measured velocities of 1 to 5 km/s for G-band bright
points in the ``moat'' around a sunspot; they showed that the motions
are constrained to the intergranular lanes and are primarily driven by
the evolution of the granulation pattern. They found that the bright
points continually split and merge, with a mean time between splitting
events of few hundred seconds. Berger et al. (1998) observed a
similar rapid splitting and merging of bright points in an enhanced
network region.

In the present paper we use observational data from the SVST on La
Palma to derive time-dependent granulation flow fields and to
simulate the horizontal motions of magnetic elements in the
photosphere. The data were collected on 1995 October 5 between
10:57 and 12:08 UT. Observations were made simultaneously with
two CCD cameras: one used a 12 {\AA} bandpass interference filter
with a center wavelength of 4305 {\AA} (G band), and the other
used a 54 {\AA} bandpass filter with a center wavelength of
4686 {\AA}. Frame selection was based on the G band images:
only the three best frames in each 20 s evaluation period were
retained for analysis. Both cameras were equipped with phase-diversity
beam splitters which put two images on each CCD with a difference in
focus position. The images were corrected for seeing effects using
Partitioned Phase-Diverse Speckle restoration, and were carefully
coaligned using image destretching techniques. The data were
space-time filtered to remove the effects of solar p-mode
oscillations. The result is a time series of 180 images with very high
spatial resolution covering a period of about 70 min. The image scale
is 0.083 arcsec/pixel and the mean time between frames $\Delta t$ =
23.5 s.

The field-of-view (FOV) is a 29 by 29 arcsec area near solar disk
center containing an enhanced network region. This region is the
same as the ``network FOV'' described by L\"{o}fdahl et al. (1998),
and we refer to their paper for a detailed description the observation
and image restoration procedures. The data used here are in essence an
early version of these data (L\"{o}fdahl 1996).

Berger et al. (1998) analyzed ``the network FOV'', where magnetic
elements are seen as bright points with high contrast in the G band
(which includes the molecular bandhead of CH) and with reduced
contrast in the 4686 {\AA} band (which contains continuum and
many absorption lines). The solar granulation shows up with nearly
equal contrast in both types of images. The bright points are
generally located in the dark intergranular lanes. Subtraction of the
G band and 4686 {\AA} images yields a difference image which shows
the bright points and surrounding diffuse emission with unprecedented
clarity. Berger et al. identified bright points using a threshold
technique applied to these difference images. They define a ``magnetic
region'' as the area within their FOV which is covered by bright points
at any time during the movie. Using local correlation tracking (LCT)
with subfields of 0.4 arcsec, they measure a mean granulation flow
velocity of 0.641 km/s inside this magnetic region and 0.997 km/s in the
surrounding quiet region. They also followed the motions of bright
points using an object tracking technique, and found that the bright
points have a broad velocity distribution which peaks at 0.1 km/s but
extends to several km/s; the mean velocity is 0.815 km/s. The bright
points continually split up and merge; the average time between
merging and splitting events is 220 s. Some objects can be followed
for the entire 70 min duration of the movie.

Following Berger et al. (1998), we construct a time series of
``magnetic'' difference images by subtracting the 4686 {\AA} continuum
images from the corresponding G-band images:
\begin{equation}
I_{magn} (x,y,t) = I_{4305} (x,y,t) - I_{4686} (x,y,t) ,
\end{equation}
where $x$ and $y$ are horizontal coordinates on the Sun, $t$ is the
time, and $I_4305$ and $I_4686$ are intensities normalized to the mean
intensity in non-magnetic areas of the frame. Examples of such
difference images are shown in Figures \ref{magn_image}a
and \ref{magn_image}b (frames 40 and 177 of the time sequence).
Note that the bright points form linear structures which fill the
intergranular lanes and sometimes surround the granules on all sides.
We also construct a time series of magnetic ``masks'' by
smoothing the $I_{magn}$ images in space and time (running average
over 5 frames) and applying a $2 \sigma$ threshold. These masks define
the general areas where the bright points are located.

\placefigure{magn_image}

\section{Measurement of Bright Point Motion}

One of the difficulties in tracking bright points is that they often
have complex shapes, so that their positions are not always
well defined. Moreover, the bright features in G-band images
continually split up and merge, causing sudden changes in position
as determined using object tracking techniques (Berger {\it et
al.~}1998). To overcome these problems, we developed a new tracking
technique which uses small but finite-size ``corks'' as tracers of
bright points. The corks are small circular disks which move about on
the photospheric plane but are not allowed to overlap each other. Each
bright feature in the magnetic image is associated with a cluster
of such corks. The corks are advected by an artificial flow field
${\bf v} (x,y,t)$ which drives the corks to the brightest regions in
the magnetic image (${\bf v} \propto \nabla I_{magn}$). Clusters of
corks collect at the bright points and follow the bright points as the
brightness pattern evolves. The advantage of this method is that the
shapes of the cork clusters can adjust themselves to the actual shape
of the bright points.

In the present study we used 1400 corks with radii $r_0$ = 60 km,
equal to the pixel size of the data set. The positions of the corks
are advanced with a time step equal to one half of the average
time between frames in the movie, but the results are saved only
for whole frames ($\Delta t = 23.5$ s). The correction
for cork overlap is performed iteratively at each time step: we first
search for partially overlapping corks (i.e., pairs of corks with
separations less than $2r_0$) and for each overlapping pair we correct
the positions by moving the two corks apart in opposite directions
along their separation line until the corks just touch each other.
We then repeat the search to check whether any new overlapping pairs
occurred as a result of the previous corrections, and if necessary we
correct the positions again. The process is repeated until there are
no overlapping corks before proceeding to the next time step.
The initial positions of the corks are chosen at random from within
the magnetic mask for frame 3 (the first frame for which such a
time-averaged mask can be computed) and we follow the corks until
frame 177.

The intensity of the bright points varies with time, and some bright
points fade away, leaving the associated corks with nowhere to go.
Therefore, on every fifth frame we remove corks which fall outside
the magnetic mask (see \S 2) and reinsert new corks at other random
locations within the mask. Typically about 5\% of the 1400 corks are
replaced every fifth frame, and the total number of corks present
over the entire time interval is 3827. Hence, the corks have lifetimes
$N_k$ which are multiples of 5 frames and range from 5 to 175 frames
(here $k$ is the cork index: $k = 1, \cdots , 3827$). In this way we
obtain cork positions $(x_{k,n},y_{k,n})$ and velocities $(v_{x,k,n},
v_{y,k,n})$ as function of time for each cork, where $n$ is the time
index ($n = 1, \cdots , N_k$). These cork velocities represent
measurements of the velocities of the bright points.

Figures \ref{magn_corks}a and \ref{magn_corks}b show the spatial
distributions of corks for frames 40 and 177 plotted over
the corresponding granulation image. Note that the corks form
dense clusters located in the intergranular lanes. Comparison with
Figures \ref{magn_image}a and \ref{magn_image}b shows that the spatial
distribution of the corks closely resembles the distribution of
the bright points.

\placefigure{magn_corks}

The random motions of the photospheric flux tubes produce MHD waves
and other disturbances that propagate into the upper solar atmosphere
and contribute to chromospheric and coronal heating. Models of wave
generation (e.g. Choudhuri, Dikpati \& Banerjee 1993) predict that
short-period motions are much more effective in generating waves
than long-period motions. Therefore, it is of interest to determine
the velocity autocorrelation function which describes how the
velocity of a bright point varies with time. The autocorrelation
of velocity $v_{x,k,n}$ over a time delay $t_m = m \Delta t$
is given by:
\begin{equation}
C_{x,k,m} \equiv \frac{1}{N_k-m} \sum_{n=1}^{N_k-m} v_{x,k,n}
~ v_{x,k,n+m} ,
\end{equation}
and similar for $C_{y,k,m}$. By averaging these functions for
all corks with lifetimes $N_k > m$, we obtain mean autocorrelations
$\bar{C}_{x,m}$ and $\bar{C}_{y,m}$.

Figure \ref{magn_auto} shows the mean autocorrelations $\bar{C}_{x,m}$
and $\bar{C}_{y,m}$ as functions of delay time $t_m$ for $t_m < 1000$ s.
Note that the autocorrelation decreases monotonically with time, i.e.,
there is no evidence for any specific periodicities in the bright
point motion. The autocorrelation decreases rapidly with time over
the first few frames and more slowly thereafter. The initial decrease
is probably due to measurement noise, but the more gradual decrease on
a time scale of a few hundred seconds is probably real. We find that
the data can be fit with the following function:
\begin{equation}
C_m = 0.09 e^{-m^2} + \frac{0.12}{1 + ( t_m / \tau_0 )^2 }
+ 0.01 ~~~~~~~ [ {\rm km^2 ~ s^{-2}} ] , \label{eq:cor}
\end{equation}
where the first term represents the measurement noise, the second term
describes random motions with a correlation time $\tau_0$ = 100 s, and
the last term describes motions with a correlation times exceeding
1000 s. The effectiveness of the random motions in generating of flux
tube waves will be discussed \S 6.

\placefigure{magn_auto}

The measurement errors represented by the first term in equation
(\ref{eq:cor}) are probably due to telescope guiding errors and
residual seeing effects. To reduce these effect, we computed mean
cork velocities averaged over blocks of 5 frames (118 s).
The average of these mean velocities is $v_{x,avg}$ = -57 m/s and
$v_{y,avg}$ = +13 m/s, and the standard deviation is $v_{x,rms}$
= 379 m/s and $v_{y,rms}$ = 385 m/s. Note that the averaging time
is comparable to the time scale of the random motions, therefore
the above values likely underestimate the true velocities on the Sun.
Clearly, to measure bright point velocities more accurately, improved
observations with shorter sampling times ($\sim 10$ s) are required.

\section{Simulation of Granulation Flow and Flux Tube Motion}

Models of the interaction of magnetic fields with convection show that
the magnetic flux tubes are expelled by the granulation flow
(e.g. Schmidt et al. 1985). Convection models without magnetic
field indicate that at depths greater than a few 100 km the
convection consists of cool plumes of rapidly downflowing plasma
embedded in a connected region of warm gentle upflow (Stein \&
Nordlund 1989, 1994; Chan \& Sofia 1986; Cattaneo et al. 1991;
Rast et al. 1993). The plumes are usually located below the
vertices of the
intergranular lanes visible at the solar surface. Magneto-convection
simulations predict that the magnetic field is swept into these
downflow plumes (Nordlund \& Stein 1989, 1990; Stein, Brandenbrug \&
Nordlund 1992; Nordlund et al. 1992). In weak field regions (average
magnetic flux density $< 50$ G) the magnetic field will likely be
concentrated into discrete vertical flux tubes located within the
downflow plumes, consistent with observations of network bright
points at vertices of intergranular lanes. On the other hand, in
strong field regions ($B \sim 500$ G) the magnetic field may fill the
intergranular lanes, acting as rigid walls which make the connected
downflows extend much deeper than in the field-free case (Stein,
Brandenburg \& Nordlund 1992).

The photospheric flux elements are driven by granulation flows which
push the elements into the intergranular lanes. Berger {\it et
al. }(1998) used local correlation tracking (LCT) to derive horizontal
flow fields from a time series of granulation intensity images.
A disadvantage of this LCT technique is that it is insensitive
to stationary flows, and therefore cannot detect the more or less
steady flow of plasma from the centers of granules to the
intergranular lanes. LCT can detect such flows only if they are
very fragmented and the spatial resolution is sufficient
to resolve the fragments as they are advected from the center of the
granule to the boundary. In this paper we use a different technique
which does not require the presence of such fine-scale structures.

We derive a model of the granulation flow velocity ${\bf v} \equiv
[v_x (x,y,t), v_y (x,y,t)]$ in the photosphere. The model is based
on the assumption that the vertical component of vorticity is small,
so that the velocity can be approximated by
\begin{equation}
{\bf v} = - \nabla \phi , \label{eq:grad}
\end{equation}
where $\phi (x,y,t)$ is the flow potential. We assume that $\phi
(x,y,t)$ is well correlated with the granulation intensity: $\phi$
is positive in the bright granules and negative in the dark
intergranular lanes, hence $- \nabla \phi$ represents the radial
outflow from the cell centers to the lanes. The granulation intensity
$I_{conv} (x,y,t)$ is defined by removing the bright points from
the 4686 {\AA} image as much as possible; this is done by subtracting
a multiple of the ``magnetic'' image from the 4686 {\AA} image:
\begin{equation}
I_{conv} (x,y,t) = I_{4686} (x,y,t) - b_0 I_{magn} (x,y,t) ,
\end{equation}
where $b_0 \approx 0.5$. We assume that the acceleration of the
photospheric plasma is given by:
\begin{equation}
\frac{\partial {\bf v}} {\partial t} + {\bf v} \cdot \nabla {\bf v} =
- c_1 \nabla I_{conv} + \nu \nabla^2 {\bf v} - \frac{\bf v}{\tau_f} ,
\end{equation}
where the first term on the RHS describes the driving of the flow
by horizontal temperature gradients (the temperature is assumed
to be proportional to $I_{conv}$), the second term describes the
effects of turbulent viscosity ($\nu$ = 70 ${\rm km^2/s}$), and
the third ``frictional'' term describes the decay of the velocity
of a granule after the granule has disappeared from the temperature
field (decay time $\tau_f$ = 300 s). Inserting equation
(\ref{eq:grad}) we obtain:
\begin{equation}
\frac{\partial \phi} {\partial t} = c_1 I_{conv} +
\frac{1}{2} | \nabla \phi |^2  + \nu \nabla^2 \phi -
\frac{\phi}{\tau_f} + {\rm constant} ,
\end{equation}
where the constant is chosen such that the average of $\phi$ over the
image vanishes at all times.
The proportionality constant $c_1$ is determined by comparing the
computed flow velocities with those found in the 3D granulation
simulations of Stein \& Nordlund (1994). At a depth of 40 km
in Stein \& Nordlund's model the irrotational component of horizontal
velocity (i.e., after removal of vortical flows) has an rms value
$v_{x,rms} \approx$ $v_{y,rms} \approx$ 1.60 km/s. We adjusted $c_1$
to obtain essentially the same value in the present model.

Figure \ref{sim_vect} shows the velocity in a subarea of frame 40 as
computed with the above 2D model. Note that the velocity changes
rapidly across the intergranular lanes. Figure \ref{sim_div}a shows
the divergence of the velocity for the full frame. For comparison,
Figure \ref{sim_div}b shows the divergence of horizontal velocity at a
depth of 40 km as derived from the 3D simulations of Stein \&
Nordlund (1994). Note that the intergranular lanes are very narrow in
both cases. In our 2D model the width $w$ of the lanes is controlled
by the ``viscosity'' parameter $\nu$, and is given by $w \approx \nu
/ v_0$, where $v_0$ is the velocity of the converging flows on
either side of the lane (typically $v_0 \sim 2$ km/s). With $\nu$ =
70 ${\rm km^2/s}$ we predict $w \approx 35$ km, but due to numerical
diffusion the lanes in our model are actually somewhat wider,
$w \approx 100$ km.

\placefigure{sim_vect}

We use the above granulation flow field to simulate the horizontal
motions of magnetic elements in the photosphere. The
magnetic field is modeled as a collection of cylindrical flux tubes
which stand vertically in the solar atmosphere and each have a radius
$r_0$ = 60 km. Each flux tube is a rigid structure which is {\it
passively advected} by the granulation flow, i.e., we neglect possible
effects of the anchoring of the flux tubes at larger depth.
The rigidity of the flux tubes simulates the fact that the
magnetic field strength $B$ cannot exceed the value for equipartition
of the magnetic pressure with the external gas pressure in the
lower photosphere. Assuming $B$ = 1500 G and $r_0$ = 60 km, each
flux tube has a magnetic flux of $1.7 \times 10^{17}$ Mx. The flux
tubes behave like rigid ``corks'' which float on the granulation
flow: converging flows produce densely packed cork clusters which
represent larger magnetic elements with shapes determined by the flow.
The motions of 1400 corks are followed from frame 3 to frame 177,
which corresponds to about 7 granulation lifetimes.
The procedure for dealing with overlapping corks is the same as
in \S 3, and the initial positions of the corks are also the same,
i.e., the corks are initially located at the positions of the
bright points. However, in the present case we do not replace corks
as they move away from the bright points; all 1400 corks keep
their identity for the full duration of the simulation.

\placefigure{sim_div}

Figure \ref{sim_corks} shows the cork positions for frames 40 and 177
superposed on the corresponding convective intensity patterns.
Note that the corks collect predominantly in clusters at the
vertices where several intergranular lanes intersect. Linear
structures extend from these clusters out into the intergranular
lanes. This is due to the narrow width of the velocity lanes,
which prevents the corks from all collecting at the vertex points.
In contrast, in models with broader lanes (not shown) most of the
magnetic elements collect in a few large clusters. Although the
present model does not reproduce the exact locations of all
bright points in the final frame, the qualitative agreement between
observed and predicted patterns is quite good (compare Figures
\ref{sim_corks} and \ref{magn_corks}). In particular, the model
shows linear structures similar to the observed filigree.

\placefigure{sim_corks}

\section{Velocity Field in the Chromosphere}

The magnetic flux tubes on the Sun expand with increasing height
in the photosphere, and above some height the neighboring flux
tubes merge to fill the available volume. This ``merging'' of the flux
tubes generally occurs in the chromosphere at a height about 1000 km
or less (depending on the horizontal distance between the flux tubes).
Since the magnetic field is nearly ``frozen into'' the plasma, the
horizontal motions of the photospheric flux tubes induce plasma flows
in the chromosphere and above. In this section, we use a
potential-field model to compute the velocity field at a height of
1500 km in the chromosphere above the observed region. 

The model is based on the assumption that the magnetic field {\it
inside} the flux tubes remains close to a potential field at all
times, i.e., we ignore the effects of field-aligned electric currents
on the magnetic structures below 1500 km height. The magnetic field is
described as a superposition of fields from discrete point sources:
\begin{equation}
{\bf B} ({\bf r},t) = \sum_k {\bf B}_k ({\bf r},t) .
\end{equation}
The sources are located at positions ${\bf r}_k (t) \equiv [x_k(t) ,
y_k(t), -d_0]$, where $d_0$ = 160 km is the depth of the sources below
the photosphere (the same for all sources). Each source is the sum of
a monopole and a (vertically oriented) dipole:
\begin{eqnarray}
B_{x,k} (x,y,z,t) & = & \frac{x-x_k(t)}{R^3}
  + 3 (a_0+d_0) \frac{(x-x_k(t))(z+d_0)}{R^5} , \\
B_{y,k} (x,y,z,t) & = & \frac{y-y_k(t)}{R^3}
  + 3 (a_0+d_0) \frac{(y-y_k(t))(z+d_0)}{R^5} , \\
B_{z,k} (x,y,z,t) & = & \frac{z-a_0}{R^3}
  + 3 (a_0+d_0) \frac{(z+d_0)^2}{R^5} ,
\end{eqnarray}
where $R \equiv [(x-x_k)^2+(y-y_k)^2+(z+d_0)^2]^{1/2}$ is the distance
to a source, $a_0$ = 1000 km is the canopy height for a single source,
and we assume that all sources have the same strength and polarity.
We define tube-like structures by tracing magnetic field lines
downward from a height of 1500 km, i.e. we ignore the ``closed''
magnetic structures in between these flux tubes. This is equivalent to
assuming that the flux tubes are embedded in a field-free gas.

The frozen-in condition implies ${\bf E} + {\bf v} \times {\bf B}
= 0$, where ${\bf v} ({\bf r},t)$ is the plasma velocity and ${\bf E}
({\bf r},t)$ is the electric field. The latter can be written as:
\begin{equation}
{\bf E} ({\bf r},t) = \sum_k {\bf E}_k ({\bf r},t) - \nabla \Phi .
\label{eq:elec}
\end{equation}
Here $\Phi ({\bf r},t)$ is the electric potential, and ${\bf E}_k
({\bf r},t)$ is the electric field due to a single source:
\begin{equation}
{\bf E}_k ({\bf r},t) \equiv - {\bf v}_k (t) \times {\bf B}_k
({\bf r},t) ,
\end{equation}
where ${\bf v}_k (t) \equiv d {\bf r}_k /dt$ is the source velocity.
The condition ${\bf E} \cdot {\bf B} = 0$ yields:
\begin{equation}
{\bf B} \cdot \nabla \Phi = \sum_k {\bf B} \cdot {\bf E}_k .
\end{equation}
The right hand side can be computed directly from the known positions
and velocities of the magnetic sources, hence the electric potential
$\Phi (x,y,h_0,t)$ at height $h_0$ = 1500 km can be determined by
numerical integration along the field lines. We use the boundary
condition $\Phi (x,y,0,t)$ = 0 at the photosphere. By tracing five
neighboring field lines which are anchored in the same source, we
compute the partial derivatives $\partial \Phi / \partial x$ and
$\partial \Phi / \partial y$, which yields the horizontal components
of electric field, $E_x$ and $E_y$, at height $h_0$ [see equation
(\ref{eq:elec})]. The vertical component $E_z$ is found from the
condition ${\bf E} \cdot {\bf B} = 0$, and the velocity is computed
from ${\bf v}$ = ${\bf E} \times {\bf B} /B^2$. Here we assume
arbitrarily that the velocity is perpendicular to the magnetic field
lines.

To illustrate the method we first consider the case with only two
magnetic sources. The sources are located at $(x_1,y_1)$ = $(-1000,0)$
km and $(x_2,y_2)$ = $(+1000,0)$ km, so that there is a single
magnetic null point located at a height $z_0$ = 329 km above the
midpoint between the two sources ($x = y = 0$). Figure \ref{two_lines}
(PLATE) shows a perspective view of the magnetic field lines for this
case. Note that the field lines in $x < 0$ are connected to source 1
while those in $x > 0$ are connected to source 2; the plane $x = 0$
is the {\it separatrix surface} between the two flux tubes. The field
lines within this separatrix surface fan out more or less radially from
the null point, i.e., the separatrix surface is a {\it fan plane}.
Field lines which approach the null point along the $x$ direction are
diverted in all directions parallel to the fan plane.

Now assume that source 1 is moving with velocity 1 km/s in the $-y$
direction, and source 2 is moving with the same velocity in the $+y$
direction, i.e. perpendicular to the line connecting the two sources.
Figure \ref{two_vel} shows the predicted chromospheric velocity at
height 1500 km for this case. Note that the velocities directly above
the sources are indeed about 1 km/s, but as we approach the separatrix
surface ($x = 0$) the magnitude of the velocity increases
dramatically. Apparently, the movement of the flux tubes past each
other produces large velocities near the separatrix plane.
Simulations with higher resolution indicate that $| v_y |$ increases
without bound as $x \rightarrow 0$. This effect is due to the 3D
geometry of the magnetic field near the null point: the motion of
the magnetic sources causes field lines to sweep past the null point
in the $y$ direction, producing rapid motion of field lines in the
vicinity of the fan plane. In Appendix A we show that the functional
dependence of the velocity near the separatrix surface is determined
by the eigenvalues of the magnetic gradient tensor at the null point.

\placefigure{two_vel}

In the present model we assume for simplicity that $\Phi (x,y,0,t)$ =
0, which implies that {\it the flux tubes exhibit no rotational motion
about their vertical axes}. As two neighboring flux tube move relative
to each other, the line connecting the two flux tubes rotates with
time, but this rotation is not compensated by a corresponding rotation
of the individual flux tubes. This leads to enhanced velocities at the
separatrix surface as discussed above. Of course, if the flux tubes
were to rotate at the same rate as their relative motion, then the
chromospheric velocity enhancement at the separatrix plane would not
occur. Hence, if there are only two flux tubes, it is possible to
avoid such velocity enhancements by a suitable choice of rotation rate
of the individual flux tubes. However, on the Sun there are {\it
multiple} flux tubes which randomly move among each other. Each flux
tube has several neighbors, and the different lines connecting a flux tube
with its neighbors generally rotate at different rates. Therefore,
in general it is not possible to eliminate all chromospheric velocity
enhancements by suitable choice of the flux tube's rotation rate.

We now use the above method to predict the magnetic field and velocity
field above the observed network region. The positions of the magnetic
sources are taken from the simulation of \S 4. Figure \ref{sim_lines}
(PLATE) shows a perspective view of field lines for a small area in
frame 40. The horizontal size of the region is 3000 km, and its
location within frame 40 is indicated by the white box in Figure
\ref{sim_corks}a. Note that each field line can be traced to a
particular cork at $z = 0$.

Figure \ref{sim_vel}a shows the horizontal component of velocity in
the chromosphere ($h_0$ = 1500 km) as computed with the above method.
Figure \ref{sim_vel}b shows the boundaries between different flux
tubes at this height, i.e., each domain is connected to one particular
``cork'' at $z = 0$. Note that the predicted chromospheric velocities
are several km/s, much larger than the assumed velocities of the
magnetic sources below the photosphere. This amplification is partly
due to the spreading of the flux tubes with height, which enhances
the rotational motion of neighboring flux tubes around each other
(van Ballegooijen 1986). However, Figure \ref{sim_vel}b shows that
the largest velocities occur near separatrix surfaces where
neighboring flux tubes slide past each other. Therefore, most of
the effect is due to the magnetic null points, which enhance the
velocities of field lines which pass in their vicinity. In fact,
in some regions the field-line velocities are so large that the plasma
inertia cannot be neglected and the potential-field model breaks down.
Clearly, to obtain an accurate description of such flows a more
realistic MHD model of interacting flux tubes is needed.

\placefigure{sim_vel}

\section{Discussion}

The simulations in \S 4 show that the observed velocities and spatial
distribution of the G-band bright points are consistent with passive
advection by the granulation flow, i.e., we find no evidence for
deep-seated flows different from the observed granulation pattern.
To reproduce the observed ``filigree'' structures, we had to take into
account that the photospheric flux elements are nearly incompressible
($B \sim 1500$ G) and that the intergranular lanes are very narrow
($w \sim 100$ km). In contrast, in models with broader lanes most of
the magnetic elements collect in a few large clusters at the vertices
where several lanes intersect. Therefore, the observed filigree
structures provide indirect evidence for the existence of narrow
intergranular lanes with widths $\sim 100$ km. Note that the present
granulation flow model is based on the {\it observed} granulation
pattern and therefore includes the effects of the magnetic field on
the granulation flow (longer lifetime of granules; reduced cell size).

Three-dimensional models of solar convection (e.g. Stein \& Nordlund
1994) show the presence of small-scale vortical flows within the
intergranular lanes on scales $\sim 100$ km. Such small-scale flows
likely play an important role in breaking up the ``filigree'' into
distinct magnetic elements, as is observed in high-resolution G-band
images. Indeed, bright points sometimes exhibit rotational motions
about each other (Berger \& Title 1996). In the present paper we
neglected the vorticity of the granulation flow and its effects on
the dynamics of flux tubes. The reason is that the observational data
do not provide information about the vorticity of granulation flows
on small spatial scales. Using correlation tracking, it is possible
to measure vorticity on length scales of several granules (e.g. Wang
et al. 1995), but such large-scale vorticity probably has little
effect on the small-scale structure and dynamics of magnetic elements
within the intergranular lanes. Clearly, future modeling of flux tube
dynamics will need to take vortical flows into account, and work along
these lines is in progress (Nordlund \& Stein, in preparation).
To measure small-scale vortical flows directly (other than by tracking
magnetic elements) will require observations with much higher spatial
resolution than are presently available. 

How effective are the observed motion in generating MHD waves?
In \S 3 we show that the autocorrelation function of the bright
point velocity is approximately given by:
\begin{equation}
C( \tau ) \equiv < v_x (t) v_x (t+ \tau ) > = \frac{\sigma^2}
{1 + ( \tau / \tau_0 )^2 } ,
\end{equation}
where $\sigma^2 \approx 0.12 ~ {\rm km^2 s^{-2}}$ and $\tau_0 \approx
100$ s. The velocity power spectrum is obtained by taking the Fourier
transform of the autocorrelation function:
\begin{equation}
P( \omega ) \equiv \frac{1}{2 \pi} \int_{- \infty}^{+ \infty}
C( \tau ) e^{i \omega \tau} d \tau = \frac{1}{2} \sigma^2 \tau_0
e^{- \tau_0 | \omega |} .
\end{equation}
Hence, most of the power is contained in frequencies $\omega <
\tau_0^{-1}$ $\approx 0.01$ rad/s. This is significantly less than
the acoustic cutoff frequency in the photosphere, which is given
by $\omega_{ac}$ = $c/(2H) \approx 0.03$ rad/s, where $c \approx 7$
km/s is the sound speed and $H \approx 115$ km is the pressure
scale height. Therefore, it is doubtful that the observed motions
are very effective in generating longitudinal tubes waves, which
are evanescent at these frequencies.

A number of authors have proposed that the corona is heated by
{\it transverse} tube waves (Hollweg 1984; Choudhuri, Auffret \&
Priest 1993; Choudhuri, Dikpati \& Banerjee 1993). Muller {\it et
al.~}(1994) measure a mean speed of network bright points of 1.4 km/s
and conclude that there is sufficient energy in these motions to heat
the quiet corona. However, it is not clear that the observed velocity
is entirely due to transverse waves: part of the mean velocity must be
due to long-period motions which are not effective in generating
transverse tube waves. The cutoff frequency for tranverse tube waves
is given by
\begin{equation}
\omega_c^2 = \frac{g}{8H} \frac{1}{2 \beta +1} ,
\end{equation}
where $g$ is the acceleration of gravity, $\beta = 8 \pi p_i /B^2$
is the ratio of gas pressure and magnetic pressure, and the thin flux
tube approximation has been used (Spruit 1981). Using $\beta$ = 0.3,
we find $\omega_c = 0.0135$ rad/s. If we assume that only the high
frequency modes with $| \omega | > \omega_c$ can generate transverse
waves, then the amplitude of these waves is given by:
\begin{equation}
u_t^2 \equiv \sigma^2 \tau_0 \int_{\omega_c}^{\infty} e^{- \tau_0
\omega} d \omega = \sigma^2 e^{- \tau_0 \omega_c } \approx 0.031
~~~ [ {\rm km^2 s^{-2}} ] ,
\end{equation}
which yields $u_t \approx 0.18$ km/s. The energy flux in such waves is
given by $F = \rho u_t^2 v_A$, where $\rho$ is the mass density, and
$v_A \equiv B/ \sqrt{4 \pi \rho}$ is the Alfv\'{e}n speed. Using
$\rho = 7 \times 10^{-8}$ $\rm gr/cm^3$ for the density within the
tubes in the low photosphere, we find that the energy transported by
these waves per unit magnetic flux is $F/B$ = $u_t^2 \sqrt{ \rho /
( 4 \pi )}$ = $2.3 \times 10^4$ $\rm erg / cm^2 / s / G$.
Hence, in plage regions with magnetic flux density $\sim 50$ G, we expect a
mean energy flux $\sim 10^6$ $\rm erg / cm^2 / s$. This is
comparable to the observed radiative and conductive losses from
the corona above plage regions (Withbroe and Noyes 1977).

Extrapolations of magnetic and velocity fields to larger heights
(see \S 5) show that the chromospheric velocity can be locally much
larger than the horizontal velocities of the flux tubes in the
photosphere. This enhancement is due to the 3D geometry of the flux
tubes and the dynamics of field lines near magnetic null points
where different flux tubes interact. This result may have important
implications for models of coronal heating. First, the coronal heating
rate varies as the {\it square} of the footpoint velocity, hence the
velocity enhancement found here could significantly increase the
coronal heating rate. Second, the enhanced velocities are localized
near separatrix surfaces, and therefore could affect the way in which
electric currents build up at such surfaces (Glencross 1975, 1980;
D\'{e}moulin \& Priest 1997). Finally, the predicted chromospheric
velocities are so large that plasma inertia cannot be neglected;
quasi-static potential-field models cannot accurately describe the
flows that occur where neighboring flux tubes interact. Clearly,
to understand how magnetic energy is stored and dissipated in the
corona, a more realistic {\it dynamical} model of interacting flux
tubes is needed.

\acknowledgements
The observational data used in this study were kindly provided by
R. A. Shine of Lockheed-Martin, and were obtained at the SVST, which
is operated by the Swedish Royal Academy of Sciences at the Spanish
Observatorio del Roque de los Muchados of the Instituto de
Astrophys\'{i}ca de Canarias (IAC). We thank T. E. Berger for many
useful discussions regarding his analysis of these data. This work
was supported by NASA grant NAGW-2545 at the Center for Astrophysics.

\appendix

\section{Singularity in Velocity Near a Fan Plane}

We introduce a coordinate reference frame $(x,y,z)$ with the origin
located at a magnetic null point and the axes aligned with
the eigenvectors of the magnetic gradient tensor, $\partial B_i /
\partial x_j$. Then, apart from a constant factor, the potential
field near the null point can be written as:
\begin{equation}
B_x = - x, ~~~~ B_y = a y, ~~~~ B_z = (1-a) z ,
\end{equation}
where $a$ is a dimensionless parameter describing the ratio of
$x$ and $y$ eigenvalues. We assume that $a$ lies in the range
$0 < a < 1$, so that $x = 0$ is the fan plane. Now suppose that the
plasma velocity is constant in time and is parallel to the fan plane:
${\bf v} = v_y (x) \hat{\bf y}$. The induction equation $\nabla
\times ( {\bf v} \times {\bf B} ) = 0$ yields $dv_y /dx$ =
$- a v_y(x)/x$, which has the solution $v_y (x) \propto x^{-a}$.
Since $a > 0$, the velocity $| v_y |$ increases without bound as we
approach the fan plane ($x \rightarrow 0$). This singularity is not
present in the two-dimensional case ($a = 0$).

%
%

\clearpage

\figcaption[f01.eps]{
Difference images obtained by subtracting G-band and continuum images
obtained at the Swedish Solar Observatory at La Palma on 1995 October 5:
(a) frame 40; (b) frame 177.
\label{magn_image}}

\figcaption[f02.eps]{
Tracking bright points with finite-size corks: (a) frame 40;
(b) frame 177. The image shows the corrected granulation intensity,
$I_{conv} (x,y)$.
\label{magn_corks}}

\figcaption[f03.eps]{
Autocorrelation function of the measure bright point velocity.
\label{magn_auto}}

\figcaption[f04.eps]{
Simulated granulation flow velocity in a subarea of frame 40.
\label{sim_vect}}

\figcaption[f05.eps]{
Divergence of granulation flow velocity: (a) as computed for frame 40
of the La Palma data using the 2D model presented in \S 4; (b) as
derived from 3D simulations of solar granulation by Stein \& Nordlund
(1994), displayed on the same the spatial scale and resolution as the
La Palma data. Note that the intergranular lanes are very narrow in
both cases.
\label{sim_div}}

\figcaption[f06.eps]{
Cork positions for (a) frame 40, (b) frame 177.
\label{sim_corks}}

\figcaption[f07.eps]{
Potential-field model of two interacting magnetic flux tubes in the
solar atmosphere.
\label{two_lines}}

\figcaption[f08.eps]{
Horizontal velocity in the chromosphere above two magnetic sources.
\label{two_vel}}

\figcaption[f09.eps]{
Potential-field model of magnetic flux tubes in the solar atmosphere.
The model corresponds to a small region in frame 40 of the La Palma
data set (the region is indicated by the white box in
Fig. \ref{sim_corks}a).
\label{sim_lines}}

\figcaption[f10.eps]{
(a) Horizontal velocity at a height of 1500 km in the chromosphere
as predicted from the potential-field model for the region shown
in Fig. \ref{sim_lines}. The longest arrows in this plot correspond
to 5 km/s, and the bold arrows show the direction of the field in
regions where the horizontal velocity exceeds 5 km/s.
(b) Separatrices between different flux tubes.
\label{sim_vel}}

\end{document}